# Generating strong magnetic flux shielding regions in a single crystal of $Bi_2Sr_2CaCu_2O_8$ using a blind hole array

Gorky Shaw[1], Biplab Bag[1], S. S. Banerjee[1,*], Hermann Suderow[2] and T. Tamegai[3]

[1]Department of Physics, Indian Institute of Technology, Kanpur-208016, India
[2]Laboratorio de Bajas Temperaturas, Departamento de Física de la Materia Condensada, Instituto de Ciencia de Materiales Nicolás Cabrera, Facultad de Ciencias Universidad Autónoma de Madrid, E-28049 Madrid, Spain.
[3]Department of Applied Physics, The University of Tokyo, Hongo, Bunkyo-ku, Tokyo 113-8656, Japan.

*E-mail: satyajit@iitk.ac.in

**Abstract:** Magneto-optical imaging studies in a single crystal of $Bi_2Sr_2CaCu_2O_8$ partially patterned with a hexagonal array of pinning centers (blind holes) reveals local features in the patterned region which are distinct compared to the pristine unpatterned regions in the sample. The patterned area exhibits a strongly diamagnetic local magnetization response and is characterized by a local penetration field enhanced by a factor of three. We show that strong shielding currents around the periphery of the nanopatterned region create a barrier which prevents vortex entry into the patterned region thus sustaining an effectively flux-free state upto the enhanced penetration field.

# 1. Introduction

Controlling the strength and location of pinning centers in superconductors is important for understanding the different glassy phases of pinned vortex state. From the application point of view, pinning of vortices is of paramount importance as it prevents vortex flow related dissipation inside the superconductor when current is sent through the sample. Pinning of vortices imparts to the superconductor an important property viz., the critical current density of the superconductor ($J_c$), viz., the threshold current value below which the superconductor can carry current without producing dissipation. A variety of different pinning centers have been studied for pinning vortices. For random point pinning centers, long-range order is lost and short-range order exists in the vortex lattice wherein the vortex correlations extend over few intervortex lattice spacings ($a_0 \propto \sqrt{\phi_0 / B}$ where $\phi_0 = 2.07 \times 10^{-7}$ G-cm$^2$ is the magnetic flux quantum of a single vortex and $B$ is the magnetic field) [1].When pinning is weak, the pinning force and hence the critical current density ($J_c$) is inversely related to the size of region within which the vortices are correlated. The scope for enhancement in critical current by chemically doping the superconductor is limited by the choice of dopants and the changes produced in the materials properties. With heavy ion irradiation induced damage along the tracks of passage of ions through the superconducting material, one can produce randomly distributed extended defects called columnar defects. Due to effective localization of the vortices on the columnar defects, it is found that columnar defects produce much higher enhancements in $J_c$ as compared to superconductors with random point pinning centers [1-3]. Point pins and columnar defects are both random pinning centers, and it is interesting to characterize the effect of more controlled ways to generate pinning centers using regular ordered patterns. Modern microfabrication and nanoengineering techniques such as electron beam lithography and Focused Ion Beam have made it possible to generate arrays of artificial pinning centers in thin films as well as single crystals of superconductors. Distance between pinning centers and their sizes can be tuned to be comparable to the superconductor parameters like the superconducting coherence length ($\xi$) or the penetration depth ($\lambda$). By patterning superconductors down to mesoscopic length scales a variety of novel exotic phenomena have emerged, viz., formation of giant vortex states [4], non-integer magnetic flux in Al thin films [5], and significant enhancements in pinning found in superconducting samples with a correlated array of magnetic nanodots [6]. Presence of periodic pinning centers leads to commensurable effects wherein there is an enhancement in pinning when the density of vortices is an integral multiple of the density of the artificial periodic pins. A number of studies have reported on strong pinning effects observed at commensurability in samples with periodic array of antidots (holes open at both ends in the superconductor) [6-13]. These studies have revealed features like reduction in the low-frequency noise due to vortex motion [7], reduced flux creep rate [8] and significant enhancement in the irreversible magnetization response and critical current density [8-11]. A square $2 \times 2$ array of antidots exhibits bistability features which are important from the point of view of single flux quantum based logic circuits [13]. In comparison to antidots, study of the behaviour of vortices in samples patterned with blind holes (open at only one end of the superconductor) [14-16] has received relatively less attention. We recently reported on the heterogeneous pinning properties of the vortex state created by partially nanopatterning a single crystal of 2H-NbSe$_2$ with blind holes using Focused Ion Beam (FIB) milling technique [17]. Bulk magnetization measurements on the sample nanopatterned with blind holes revealed a dynamic weak to strong pinning crossover. We also recently reported on the observation of highly non-uniform field gradients sustained in the vicinity of the patterned region in a single crystal of Bi$_2$Sr$_2$CaCu$_2$O$_8$ studied with local magneto-optical imaging technique [18].

    Antidots and antidot arrays can be effectively used to guide vortex motion [19,20]. The operation and performance of superconducting circuits are significantly affected by a magnetic field. When current is passed through the superconducting circuit element, vortices drift under the influence of the current induced Lorentz force acting on them, and produce dissipation and voltage noise, e.g. in Josephson junctions and SQUID sensors [21-23]. Therefore, in these superconducting circuits operating in a magnetic field environment, vortices become a major source of noise and dissipation in the circuit. Here we are interested in investigating ways of creating nearly flux-free regions inside the sample in the presence of applied fields which are larger than the bulk penetration field of the sample.

Vortex entry and exit from the sample edges in a superconductor is known to be affected by the presence of surface and geometrical barriers [24-26]. Presence of these barriers leads to an enhancement in the penetration field below which the sample is in a flux-free state. In principle, heavy-ion irradiation around the perimeter of a region in a sample can also make a region flux-free by pinning the penetrating vortices on the columnar defects present at the perimeter. However to ensure the above one needs to prepare a thick mask which is impervious to the irradiating ions. Such a procedure can be complicated, especially when dimensions of the region to be made flux-free are in the submicron range. In this report, we show that a periodic array of blind holes is an interesting alternative for creating flux-free region inside the sample. The close-spaced blind holes were patterned on a single crystal of the high-$T_c$ superconductor $Bi_2Sr_2CaCu_2O_8$ (BSCCO). The patterned sample was imaged for magnetic flux distribution using high sensitivity differential magneto-optical imaging (DMO) technique. By locally imaging the magnetic field distribution in the sample at low fields, we observe that the nanopatterned region with blind holes exhibits features which are distinct from the unpatterned regions of the sample. The nanopatterned area is found to exhibit strongly diamagnetic local magnetization and is effectively flux-free at fields where vortices have penetrated into the unpatterned regions of the sample. Further analysis of the MO images reveals the presence of strong shielding currents around the patterned area, comparable to large shielding currents associated with geometrical barriers present at the sample edges, which act as a barrier towards vortex entry into the patterned region. The barrier leads to gradients in vortex distribution in the sample and a significant enhancement in the effective penetration field in the patterned region.

## 2. Experimental details

*2.1. Sample Preparation*

We chose BSCCO as the material to pattern as it is a high-$T_c$ superconductor with a high $T_c$ and hence a promising material from application potential point of view. We preferred to pattern single crystals rather than thin films, as single crystals are intrinsically far weaker pinning as compared to thin films. Hence the change in pinning due to the introduction of artificial pinning centers can be more readily detected. A high quality single crystal of $Bi_2Sr_2CaCu_2O_8$ (BSCCO) [27] of dimensions (0.8 × 0.5 × 0.03 mm$^3$) and $T_c$ = 90 K was used. Focused Ion Beam (FIB) machine (dual beam FEI make Nova 600 NanoLab) was used to mill the sample surface with a focused Ga ion beam (diameter ~ 7 nm) to produce a hexagonal array of blind holes covering a rectangular area of ~ (44 μm × 39 μm) in the sample as shown in the Scanning Electron Microscopic (SEM) image in Figure 1(a). The FIB patterned region is the bright rectangle near the center of the sample, indicated by the arrow. The mean diameter of holes is 170 nm (which is of the order of the superconducting penetration depth $\lambda_{ab}$ ~ 200 nm for the *ab* crystal orientation in BSCCO) and a mean center-to-center spacing between the holes (*d*) of 350 nm and depth of ~ 1 μm. For the patterned array a commensurate state is realized at the matching field value of $B_0 = 1.155\phi_0 / d^2 \approx 195$ G (where $\phi_0$ is the quantum of magnetic flux = $2.07 \times 10^{-7}$ G-cm$^2$) when $d \approx a_0$ ($\propto \sqrt{\phi_0 / B}$). Figure 1(b) shows a SEM image of the magnified portion of the patterned region, showing the well-ordered hexagonal array of blind hole pinning centers generated using FIB.

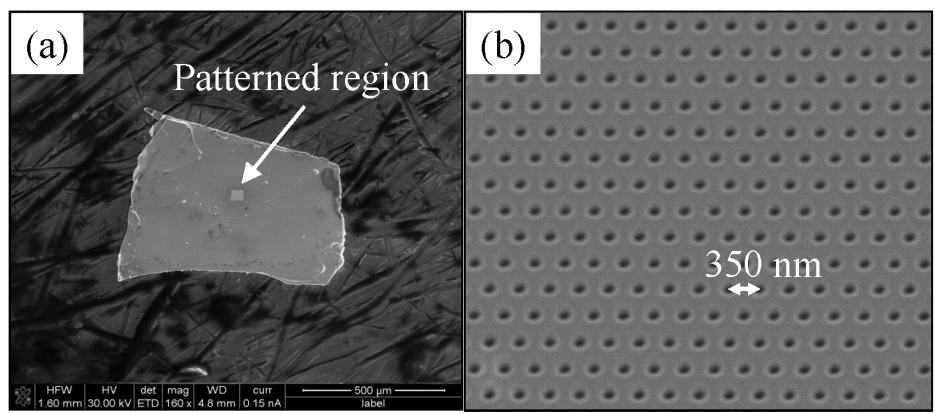

**Figure 1.** (a) Scanning electron microscope (SEM) image of the sample. The arrow indicates the location of the patterned area. (b) Magnified portion of the patterned region, showing the well-ordered hexagonal array of blind holes.

*2.2. Magneto-optical Imaging*

Conventional magneto-optical imaging (MOI) [28,29] and differential magneto-optical imaging (DMO) [30-32] techniques were employed to image the changes in magnetic flux distribution in and around the patterned region of the sample. In our MOI setup, at an applied magnetic field *H*, the spatial distribution of Faraday rotated light intensity, *I*(*x*, *y*), provides information about $B_z(x, y)$, the *z*-component of the local magnetic field at a location (*x*, *y*) on the sample surface ($I(x, y) \propto B_z(x, y)^2$). DMO images show changes in intensity, δ*I*(*x*, *y*), due to changes in $B_z(x, y)$ (i.e., δ$B_z$(*x*, *y*)) in the sample. The DMO images are obtained by modulating the externally applied magnetic field (from *H* to *H*+δ*H*, where δ*H* = 1 Oe) and synchronously capturing differential MO images. The differential intensity δ*I*(*x*, *y*) = <*I*(*H*+δ*H*)> – <*I*(*H*)>, where <..> represents average over *n* images (usually *n* = 20), represents δ$B_z$(*x*, *y*) in response to the external field modulation δ*H*. The magneto-optical imaging system is calibrated by carefully and repeatedly measuring under various illumination conditions, the magneto-optical intensity *I* versus *H* variation far away from the sample and determining the mean fitting parameters (like the Verdet constant). Using the fitting parameters one can determine the $B_z$ distribution across a sample from the measured MO intensity across the sample. Also, as $\delta I \propto B_z \delta B_z$, therefore by measuring the δ*I* corresponding to δ$B_z$ = δ*H* for a given *H* far away from the sample edges one determines the δ$B_z$ across the sample. Unlike in conventional DMO technique [30], due to irreversible magnetization response in the low-field regime of our measurements, we do not average the differential image by repeated modulation of the external field between *H* and *H*+δ*H*. In the differential images, bright and dark contrasts indicate high and low δ$B_z$, respectively. A schematic and further details of our MOI setup have been presented elsewhere [33]. In our measurements, the sample was cooled in nominal zero magnetic field to temperatures below $T_c$ and then *H* was increased in steps of 1.5 Oe and differential MO images were obtained by modulating the external field by δ*H* = 1 Oe at each *H*. Such measurements were performed at different temperatures. All our measurements were carried out with magnetic field applied parallel to the c-axis of the single crystal (*H*||*c*).

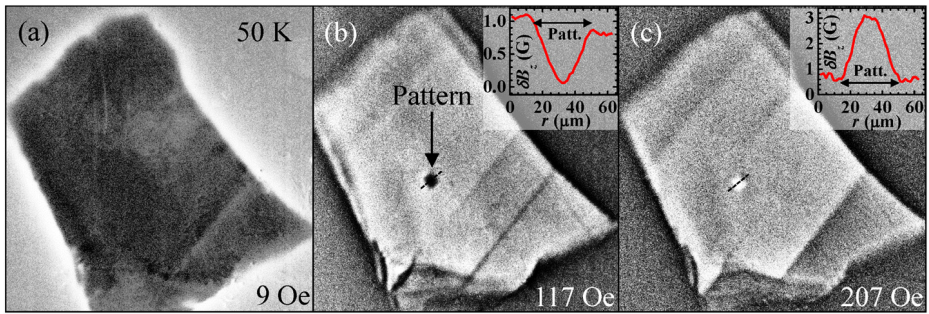

**Figure 2.** (a)–(c) Differential Magneto-Optical (DMO) images obtained at 50 K with *H* at 9, 117 and 207 Oe, respectively. The patterned region is indicated by the arrow in (b). Insets of (b) and (c) show δ$B_z$ vs. *r* (: distance along the line) behaviour along the dashed line drawn across the patterned region in the corresponding DMO images.

## 3. Results

*3.1. DMO images and evidence of magnetic field gradients in the vicinity of the patterned region*

Figure 2 shows a set of DMO images obtained at 50 K with $H$ at (a) 9, (b) 117 and (c) 207 Oe, respectively. The gray contrast outside the sample, away from the sample edges in each image, corresponds to the change in the Faraday rotated magneto-optical intensity due to the external field modulation $\delta H = 1$ Oe [32]. In fig. 2(a), we see that, at 9 Oe the sample exhibits a dark contrast, indicating $\delta B_z \ll 1$ G, i.e., significant diamagnetic shielding of changing magnetic field. This corresponds to a diamagnetic Meissner state of the superconductor. At 117 Oe (fig. 2(b)), a uniform gray contrast is seen in the unpatterned regions of the sample, indicating uniform penetration of vortices in these regions. Note that in these unpatterned regions of the sample $\delta B_z \sim 1$ G in response to a $\delta H = 1$ Oe modulation of the external magnetic field. Such a behaviour indicates reversible magnetization response of the vortex state in the unpatterned regions of the sample. However, we readily see from fig. 2(b) that in the patterned area, the DMO intensity is much lower as this region appears much darker as compared to the rest of the sample. For convenience, the patterned area has been indicated with the arrow in fig. 2(b). The dark contrast in this region indicates negligible flux change $\delta B_z$ in response to the $\delta H$ modulation at different fields. The insets of figs. 2(b) and 2(c) show the $\delta B_z(r)$ behaviour ($r$: distance along the line) along the dashed lines drawn across the patterned region as shown in the respective figures. From the line scan in fig. 2(b) it is evident that at 117 Oe, $\delta B_z$ inside the patterned region reduces to $\sim 0$ from the value of $\delta B_z \sim 1$ G at the edges due to the external field modulation $\delta H = 1$ Oe (note, as shown later, figs. 3(b) and 3(c) confirm that no flux is present inside the patterned region at 117 Oe). Such a behaviour of $\delta B_z \sim 0$ G in response to a $\delta H$ modulation at different $H$ continues upto higher $H$, viz., upto $\sim 120$ Oe. Beyond 120 Oe, regions inside the pattern area exhibit increase in $\delta B_z$ beyond $\sim 0$ G in response to $\delta H$ modulation. Although there exist channels for entry of vortices into the patterned region through the superconducting regions present in between the blind holes, yet no change in magnetic flux is observed within the patterned region with increasing $H$, until $H \geq 120$ Oe. With increase in $H$ above 120 Oe, the $\delta B_z$ inside the patterned region changes gradually with $\delta H$ modulation. At 207 Oe (fig. 2(c)), we observe the appearance of significant bright contrast over the entire patterned area, which corresponds to a $\delta B_z \sim 3$ G inside the patterned region in response to a $\delta H = 1$ Oe modulation (cf. line scan in inset of fig. 3(c)). While the patterned region appears to be strongly shielding at lower fields and largely devoid of vortices, there seems to be a significant increase in the local field $B_z$ ($\delta B_z \sim 3$ G) in response to a small change in $H$ ($\delta H = 1$Oe), allowing for a stronger concentration of enhanced flux inside the patterned region in response to $\delta H$, at higher fields. Some of our earlier observations [18] had shown dark contrast inside the patterned region, and consequent large $B_z$ gradient in the vicinity of the patterned region. The gradients in magnetic field are associated with shielding currents circulating around the patterned region. These shielding currents prevent the entry of vortices into the patterned region at lower fields. At higher $H$ it appears that the magnetic field gradients outside the patterned region weaken, allowing for an easier passage of vortices into the patterned area. Earlier reports had indicated enhanced local irreversibility induced by patterned pinning centers [15,17].

*3.2. Local diamagnetic magnetization response in the patterned region*

We deduce the local as well as bulk isothermal magnetization ($M$) vs. $H$ curves in different regions of the sample using conventional MOI, i.e., without using $H$ modulation, we use conventional MOI to capture the $B_z(x, y)$ distribution. Figure 3(b) shows one such conventional MO image, obtained at 50 K, 117 Oe. The $M$-$H$ behaviour is deduced using, $M(H) = \dfrac{\int [B_z(x,y) - H] dx dy}{4\pi A}$, where $A$ is the area over which averaging is performed and the local field $B_z(x, y)$ is determined from conventional MOI. The local $M$-$H$ corresponds to averaging over a region with area, $A = 25$ μm$^2$ situated about the center of the patterned region while $M$-$H$ for whole sample corresponds to using $A$ = entire sample area.

Figure 3(a) shows the bulk (blue, circles) and local (red, squares) *M-H* curves obtained using the above protocol. The blue arrow in fig. 3(a) identifies the location in the vicinity of which the slope of the bulk *M-H* curve changes from negative to positive thereby identifying the location of the mean penetration field $H_d \sim 40$ Oe at 50 K where vortices have penetrated into the bulk of the unpatterned regions of the sample. By noting the deviation of the *M-H* data from a linear *M* vs. *H* behaviour which is characteristic of the Meissner phase for fields $H \leq H_{c1}$ (viz., by comparing the *M(H)* data with the dashed line in fig.3), the bulk $H_{c1}$ is estimated to be $\sim 40$ Oe, and it is found to be almost the same for the patterned region also. However, we see that the overall nature of the local *M-H* response in the patterned region is significantly different from that in the patterned region. As expected from observations in fig. 2, the local magnetization curve for the patterned region shows enhanced diamagnetic response compared to the unpatterned regions of the sample. The strong diamagnetic response indicates that the patterned region is associated with enhanced shielding currents [34], which are stronger than those present in the unpatterned regions. The local *M-H* curve of the patterned region is also found to be linear upto the local $H_{c1} \sim 40$ Oe (indicated by the black arrow in fig. 3(a)). However, even beyond 40 Oe, the deviation is from linearity is gradual and the *M-H* response is strongly diamagnetic and $B_z$ is $\sim 0$ G inside the center of patterned region due to the strong shielding of the flux penetrating from the unpatterned regions of the sample (see $B_z(r)$ profile across the patterned region below the MO image in fig. 3(b)). Above the penetration field, the diamagnetic magnetization response monotonically begins to decrease with increasing *H* as the flux has reached upto the center of the superconducting region under consideration. The red arrow in fig. 3(a) identifies the penetration field $H_d \sim 150$ Oe at 50 K for the 25 μm² area inside the patterned region as compared to a value in the vicinity of 40 Oe ($\sim 45$ Oe) in the unpatterned regions of the sample. The $H_d \sim 150$ Oe of the patterned region is greater than the $H_{c1}$ of the sample. Ratio of the penetration fields indicates that the shielding currents associated with the diamagnetic response of the patterned region is at least three to four times larger than in the unpatterned regions of the sample. Figure 3(b) shows a conventional MO image obtained at 117 Oe. Note that unlike a DMO image, the intensities (bright, dark and different gray shades) in a conventional MO image represent different values of local $B_z$ (cf. discussion on MOI technique in section 2.2). It is evident from the image in fig. 3(b) that the patterned area appears much darker as compared to rest of the sample, indicating strong shielding of flux inside the patterning region.

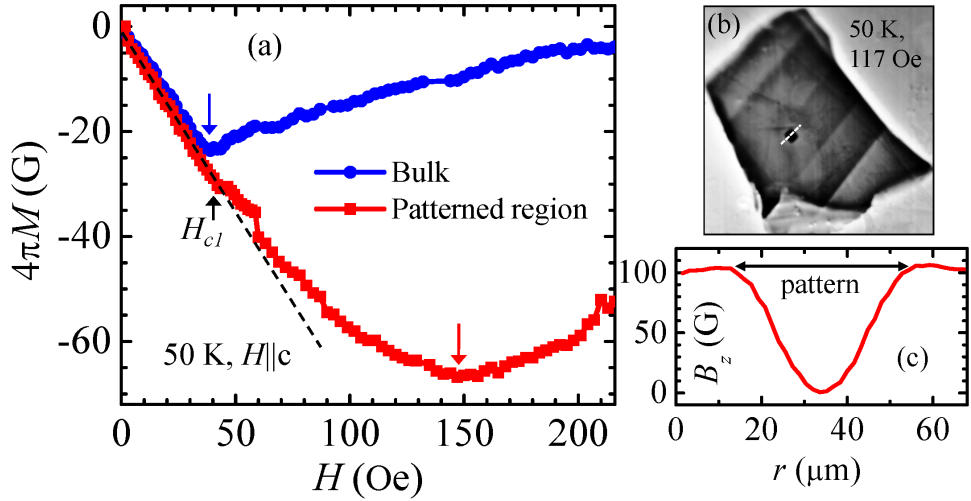

**Figure 3.** (a) *M(H)* curves obtained from MO images. The blue curve (circles) represents *M(H)* response for the entire sample area. The red curve (squares) represents local *M(H)* response in the patterned region. (b) Conventional MO image obtained at 50 K, 117 Oe (corresponding to the DMO image shown in fig. 2(b)). (c) Variation of $B_z$ along the dashed line marked in (b).

Figure 3(c) shows the variation of $B_z$ across the patterned region along the dashed line shown in fig. 3(b). Clearly, the $B_z$ reduces to ~ 0 G inside the patterned region. Enhanced shielding currents associated with the patterned region result in the stronger diamagnetic magnetization response within the patterned region compared to the bulk response, as seen in the *M-H* curves in fig. 3(a).

### 3.3. Distribution of shielding currents

We use an inversion technique [35,36] to determine the distribution of superconducting shielding current density ($j(x, y)$) across the sample area from $B_z(x, y)$ obtained from the conventional MO images an example of which is shown in fig. 3(b). We obtain the absolute value of shielding current density, $j(x,y) = |\mathbf{J}(x,y)| = |\sqrt{j_x(x,y)^2 + j_y(x,y)^2}|$, where $\mathbf{J} = j_x(x,y)\hat{x} + j_y(x,y)\hat{y}$ is the net current density at $(x, y)$ and $j_x$ and $j_y$ are the *x* and *y* components of the current density. Image in figures 4(a) and 4(d) show the ($j(x, y)$) distribution across the sample at 50 K, determined as above, at 117 Oe and 207 Oe, corresponding to the DMO images in figs. 2(b) and 2(c), respectively. In these images dark contrast indicates weak current density and bright contrast indicates regions of strong current density distribution. From both images it is evident that across unpatterned portions of the sample $j(x, y)$ is small while larger shielding currents surround the periphery of the sample as well as the patterned region. The large currents near the sample edges represent Meissner shielding currents associated with significant geometrical barrier effects in the sample. Figures 4(b) and 4(e) shows $j(x, y)$ variation along the two lines drawn across the images in figs. 4(a) and 4(d), respectively. The red and blue curves represent $j(x, y)$ across the dashed and solid lines passing over the patterned and unpatterned regions of the sample, respectively. The blue curves show that the $j(x, y)$ across unpatterned regions of the sample is small (away from the sample edges) ~ 2 - 3 × $10^4$ A/cm$^2$, while large $j(x, y)$ ~ 15 - 20 × $10^4$ A/cm$^2$ is present along the sample edges. The red curves show large $j(x, y)$ in the vicinity of the patterned region, comparable to that present along the sample edges.

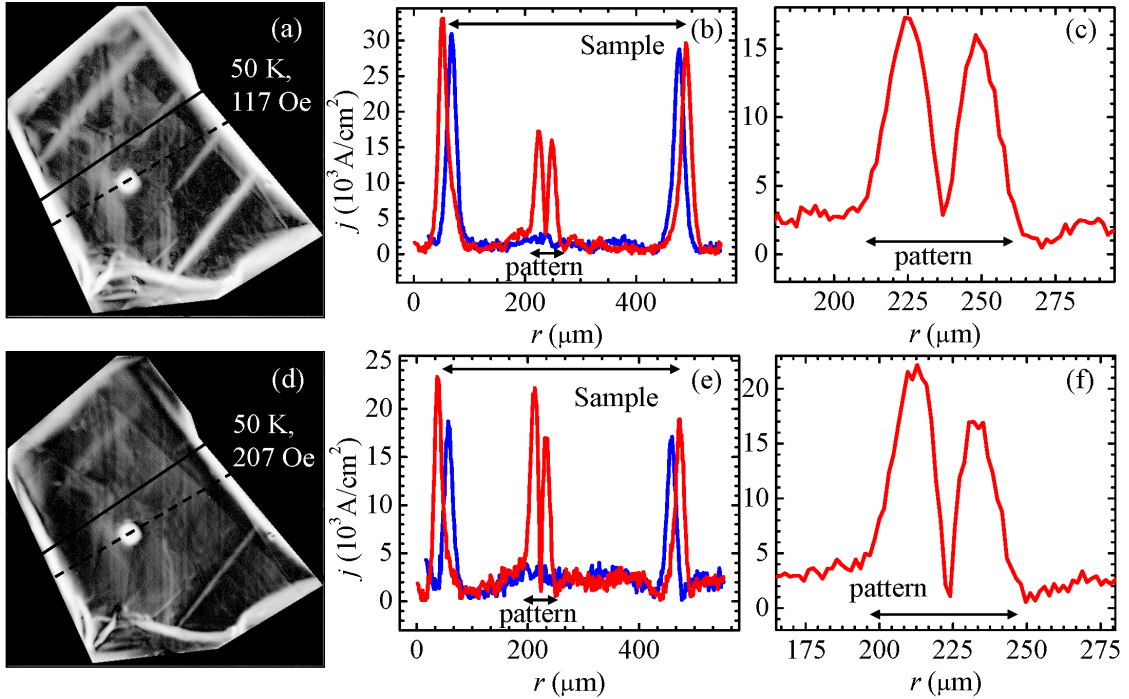

**Figure 4.** (a) and (d) Images showing $j(x, y)$ distribution across the sample at 50 K with *H* at 117 Oe and 207 Oe, respectively. (b) and (e) $j(x, y)$ vs. *r* (: distance along the line) behaviour across the two lines shown in (a) and (d), respectively. The red curves represent the dashed lines passing over the patterned region. The blue curves represent the solid lines passing over unpatterned regions of the sample. (c) and (f) Magnified portion of the red curves in (b) and (e), respectively.

This behaviour shows that the vortex distribution in and around the patterned region is more non-uniform than in the unpatterned regions of the sample. Figures 4(c) and 4(f) show a magnified portion of the red curves across the patterned region of the sample. It is clear that $j(x, y)$ present at the boundary of the patterned area is significantly large (~ 5 - 6 times larger) in comparison to that in the unpatterned regions. The presence of significant shielding currents around the patterned region creates a barrier which separates the patterned and unpatterned regions. Remarkably, around the patterned area, we get a situation analogous to surface barriers for vortex entry at the sample edges. The presence of stronger shielding currents around the patterned region in comparison to the unpatterned regions is consistent with our earlier observation of a larger penetration field for the patterned region of the sample (cf. fig. 3).

## 4. Discussion

Our experiments are performed in the limit of dilute density of vortices viz., in a low field regime wherein we approach close to the matching field (~ 195 G) at $H$ = 200 Oe. As $H$ is increased gradually from zero, vortices enter the sample from the unpatterned regions of the sample. However they are prevented from uniformly distributing inside the sample due to the presence of a patterned region inside the sample. As vortices approach the periphery of the patterned region with the blind holes they get pinned at the blind holes at the periphery and their further progression into the patterned region is strongly hindered [37]. For antidots patterned in a superconductor placed in a magnetic field, based on Ginzburg-Landau equations, it was shown [20] that encircling the periphery of each patterned hole is a shielding current distribution which decays as $exp(-r/\lambda)$ as one moves radially ($r$) outwards from the antidot. Due to physical similarities with antidots, we can assume a similar $j$ distribution around each blind hole. Thus each blind hole on the periphery of the patterned region occupied with a vortex has a shielding current $j$ circulating around it. In the dilute limit, this shielding current around each blind hold prevents other vortices from occupying the same blind hole and the vortices prefer to occupy empty blind hole sites. Thus the close-spaced patterned blind holes along the periphery of the patterned region appear to be behaving collectively with the currents circulating around each blind hole bunching together resulting in a net shielding current (cf. fig. 4). We believe that around the patterned region, the demagnetizing effects could be significantly different from the bulk of the sample (see deviation from linearity prior to the mean $H_d$ in red curve of fig. 3). Along with a distribution of penetration fields, this may result in the more rounded curvature of the $M(H)$ curve near the mean penetration field of the patterned region (cf. red curve in fig. 3) The shielding current around the periphery of the patterned region prevents the vortices from occupying the blind hole sites inside the pattered region resulting in strong gradients in magnetic field across the patterned region. Thus the shielding currents around the patterned area create a barrier towards redistribution of vortices in the patterned region.

We suggest here some differences anticipated in the behaviour between antidots and blind holes. Note while antidot is a through hole with no superconductor inside it and the hole has the same height as thickness of the superconductor, the blind hole is open only on one end and is closed at the other, with the height of the hole being significantly less than the thickness of the superconductor. The depth of the blind holes we have used is about 1 μm which is significantly less compared to the thickness of the superconductor which is ~ 30 μm. The blind hole can therefore be considered as a minor depression or dimple on the surface of the superconductor. In analogy, one may consider that an individual blind hole could be approximated as a point pinning center in comparison to the antidot which is closer to a columnar pinning center. In terms of effectiveness of individual pinning centers it is quite likely that an individual antidot could be more effective in trapping vortices as compared to an individual blind hole. However a mere modification of the intrinsic pinning strength of the material with patterning is not what we desired here as individual strong pinning centers can be achieved quite effectively in a variety of different ways like, chemically incorporating impurity and defects in a superconducting material, heavy ion irradiation, are a few common ways popularly employed, and one need not use a sophisticated technique like FIB milling at nanoscales. What we would like to emphasize here is that we have shown that one obtains a very effective enhancement in the penetration field of the superconductor by merely patterning with a well correlated array of weaker pinning blind holes. Rather than making individual strong pinning centers, we show that a correlated array of strong pinning centers leads to an effective overall enhancement in pinning characteristics. Compared to antidot patterning blind hole array over larger areas may also less resource intensive and also cost effective. The strong shielding currents around the patterned region are responsible for the strong diamagnetic magnetization response noted in the *M-H* curve in fig. 3. With increasing *H* large buildup of repulsively interacting vortices statistically enhances the attempt frequency of vortices approaching the barrier and escaping across it to enter into the patterned region, which leads to an increase in $\delta B_z$ due to a modulation of $\delta H$ by 1 Oe at high *H*, cf. DMO image for 207 Oe in fig. 2(c). It appears that by partial nanopatterning we have engineered a region within the sample within which an effectively flux-free state is sustained upto much higher fields in comparison to the unpatterned regions of the sample. The generation of a flux-free region in a superconductor can also be considered as a reversed scenario of the 'flux focusing' observed in superconducting Josephson networks or granular superconductors [38]. Note that we have used the term flux-free to signify that the penetration field in the patterned region has enhanced significantly making the region a strongly flux shielding region. Presently our technique does not possess the spatial resolution to resolve individual vortices. We would like to clarify that while referring to flux-free we do not mean a totally vortex-free region on a microscopic scale, but rather a strongly flux shielded region characterized by the absence of significant detectable magnetic flux within the blind hole patterned region.

## 5. Conclusion

To summarize, by partially patterning a superconductor with a periodic array of artificial pining centers we have created a region within the sample with features quite distinct compared to the pristine regions in the sample. The patterned area exhibits strongly diamagnetic magnetization response, indicating high local irreversibility, and is surrounded by strong shielding currents which act as a barrier towards vortex redistribution in the vicinity of the patterned region. The barrier leads to reduced vortex mobility and buildup of large gradients in vortex density in the vicinity of the patterned region which are sustained upto reasonably high fields resulting in a flux-free state being maintained within the patterned region upto a significantly enhanced local penetration field.

## Acknowledgments

SSB would like to acknowledge the funding support from Indo-Spain DST project - India, IIT Kanpur, India and CSIR-India.